\documentclass[aps,prb,reprint,showkeys,showpacs,superscriptaddress]{revtex4-1}

\usepackage{graphicx,lmodern,amsmath,amssymb,subfigure}

\begin{document}


\title{Ultrafast x-ray diffraction studies of photoexcited coherent phonons in SrRuO$_3$ thin films}



\author{D.~Schick}
\affiliation{Institut f\"ur Physik und Astronomie, Universit\"at Potsdam, Karl-Liebknecht-Str. 24-25, 14476 Potsdam, Germany}
\author{P.~Gaal}
\affiliation{Helmholtz-Zentrum Berlin f\"ur Materialien und Energie GmbH, Wilhelm-Conrad-R\"ontgen Campus, BESSY II, Albert-Einstein-Str. 15, 12489 Berlin Germany}
\author{A.~Bojahr}
\affiliation{Institut f\"ur Physik und Astronomie, Universit\"at Potsdam, Karl-Liebknecht-Str. 24-25, 14476 Potsdam, Germany}
\author{W.~Leitenberger}
\affiliation{Institut f\"ur Physik und Astronomie, Universit\"at Potsdam, Karl-Liebknecht-Str. 24-25, 14476 Potsdam, Germany}
\author{R.~Shayduk}
\affiliation{Helmholtz-Zentrum Berlin f\"ur Materialien und Energie GmbH, Wilhelm-Conrad-R\"ontgen Campus, BESSY II, Albert-Einstein-Str. 15, 12489 Berlin Germany}
\author{A.~Hertwig}
\affiliation{Bundesanstalt für Materialforschung und -pr\"ufung, Unter den Eichen 87, 12205 Berlin, Germany}
\author{I.~Vrejoiu}
\affiliation{Max-Planck-Institut f\"ur Mikrostrukturphysik, Weinberg 2, 06120 Halle, Germany}
\author{M.~Herzog}
\email{marc.herzog@uni-potsdam.de}
\affiliation{Institut f\"ur Physik und Astronomie, Universit\"at Potsdam, Karl-Liebknecht-Str. 24-25, 14476 Potsdam, Germany}\author{M.~Bargheer}
\affiliation{Institut f\"ur Physik und Astronomie, Universit\"at Potsdam, Karl-Liebknecht-Str. 24-25, 14476 Potsdam, Germany}
\affiliation{Helmholtz-Zentrum Berlin f\"ur Materialien und Energie GmbH, Wilhelm-Conrad-R\"ontgen Campus, BESSY II, Albert-Einstein-Str. 15, 12489 Berlin Germany}


\date{\today}

\begin{abstract}

We present ultrafast x-ray diffraction experiments on thin films of metallic SrRuO$_3$ (SRO) after their excitation with ultrashort intense laser pulses. Depending on the layer thickness, the data exhibit a transient splitting of the (002) SRO Bragg peak evidencing the generation and propagation of sharp acoustic strain waves. These distinct structural dynamics are due to the exceptionally fast electron-phonon relaxation that gives rise to a quasi-instantaneous thermal stress in SRO. The interpretation is corroborated by numerical simulations which show excellent agreement with the experimental findings. Despite the qualitatively different lattice dynamics for different SRO layer thicknesses, we identify a universal evolution of the transient average layer strain. The inferred discrepancy of the thermal stress profile from the excitation profile may hint toward a temperature-dependent effective Gr\"uneisen parameter of SRO.

\end{abstract}


\maketitle 


\section{Introduction}

In general, the internal energy of condensed matter is spread over various degrees of freedom such as electrons, lattice, magnetization and polarization. Intense and ultrashort laser pulses can directly excite one or more of these degrees of freedom within a given material. The subsequent dynamics of the individual subsystems and thus the entire system are then governed by the coupling strengths of the different degrees of freedom. For instance, structural dynamics of a crystalline lattice can be induced directly by infrared/THz absorption \cite{fors2011a} and Brillouin/Raman scattering \cite{bart1999,stev2002a} or indirectly by an initial electronic excitation (interband or intraband \cite{Rose1999a,hohl2000a}) and subsequent electron-phonon coupling \cite{hohl2000a,korf2007b}. Depending on the material of interest, the electronic and phononic subsystems may also be coupled to other degrees of freedom such as polarization \cite{korf2007b,dara2012a}, magnetization \cite{kime2005a,korf2008a,kiri2010a,ehrk2011a} or charge and/or orbital order \cite{ehrk2011a,lim2005a,beau2009a,ichi2011a}.
The understanding of these complex physical processes for a given material is of fundamental interest and, moreover, promises technological advances in the fields of signal processing, data storage and sensors as well as novel x-ray optics for ultrafast studies \cite{dawb2005a,eere2006a,herz2010a}. In particular, the coupling of various degrees of freedom is often mediated by the lattice. This triggered an enormous interest in time-resolved scattering techniques during the last two decades in order to monitor the laser-induced changes of the structural properties \cite{Rose1999a,barg2004b,frit2007,siwi2003a,eich2010a}.

A material of particular interest is the ``bad metal'' SrRuO$_3$ (SRO) due to its various complex physical properties such as itinerant ferromagnetism \cite{klei1996a,zies2010a}, ultrafast magnetostriction and electron-phonon coupling \cite{korf2008a,boja2012a}, negative spin polarization \cite{worl2000a}, orbital ordering \cite{jeng2006} and non-Fermi liquid behaviour \cite{kost1998,cao2008a}. This material can be epitaxially grown on single-crystal substrates (such as dielectric SrTiO$_3$ [STO]) with high structural perfection \cite{vrej2006a}. In combination with a fast electron-phonon relaxation of 200~fs or less \cite{korf2008a,boja2012a} and a very high damage threshold \cite{herz2012a}, SRO is perfectly suited as transducer material for the generation of coherent longitudinal acoustic (LA) phonons \cite{herz2012c}.

This report focuses on the structural dynamics of layered crystalline solids induced by ultrashort laser pulses. In particular, we discuss the generation and evolution of acoustic deformations of SRO thin films on a supporting STO substrate. We utilize the experimental method of ultrafast x-ray diffraction (UXRD). This technique employs the pump-probe scheme in which the laser-induced structural dynamics are probed by an ultrashort hard x-ray pulse at different time delays $\tau$ after the arrival of the excitation (pump) pulse.
After a brief introduction of the theoretical framework which describes the ultrafast build-up of laser-induced thermal stress in SRO, we present results of UXRD experiments on two thin films with thicknesses smaller and larger than the optical penetration depth of the 800~nm pump light. The transient changes in the UXRD data readily evidence a complex formation and propagation of LA phonon wavepackets. However, qualitatively different features appear for the two different samples. In particular, the thicker SRO layer exhibits a splitting of the Bragg peak as opposed to a continuous shift in case of the thinner film.
The experimental data can be precisely simulated by means of numerical models for the photoinduced structural dynamics \cite{herz2012b} and the dynamical diffraction of x-rays from these transient crystal structures \cite{herz2012a}. The universal features of the lattice dynamics are analyzed in detail by considering the spatiotemporal strain fields and the potential and kinetic energy of the photoexcited thin film as well as the substrate.


\section{Heating of SRO thin films by ultrashort laser pulses} \label{TTM}

The topic of laser-induced heating and transport properties of elemental metals (or metal layers) is a fairly well understood process \cite{klei1996a} and has been discussed many times in literature \cite{anis1974a,qiu1994a,norr2003a,chen2006a}. In the following we want to give a brief summary and apply the standard theoretical models to the case of thin films of SRO on a STO substrate.

The standard model for describing the transient processes in laser-heated metals is the two-temperature model (TTM) proposed by Anisimov \textit{et al.} \cite{anis1974a}. It assumes that the optical energy of the laser pulse is entirely absorbed by the conduction band electrons in the metal. The electronic system then rapidly thermalizes towards an elevated electron temperature via electron-electron scattering processes. The temperature of the electronic system $T_e$ then differs from the temperature of the lattice $T_l$ (phononic system) and subsequent electron-phonon collisions transfer energy from electrons to phonons until the two subsystems reach thermal equilibrium. Due to lattice anharmonicities the incoherently excited phonons produce thermal stress which eventually leads to the thermal expansion of the metal. For most metals the electron relaxation time $\tau_e$ is on the order of a few tens of femtoseconds (fs) \cite{chen2006a} and is thus shorter than the typical timescales of laser pulse durations and all other dynamical processes involved. In particular, this assumption holds for SRO which exhibits a very short electron relaxation time of $\tau_e \approx 4.2$~fs at 145~K \cite{kost1998}\footnote{At higher temperatures the electron-electron collisions become more frequent \cite{norr2003a,smith2001a} which results in an even smaller electron relaxation time at room temperature.}. This validates the consideration of an electron temperature at all times and restricts the electronic heat transport to be diffusive. Typically, the linear dimensions of the excitation and probe area on the sample surface is much larger than the penetration depth or the film thickness of the metal. Under these circumstances the differential equations of the TTM can be restricted to one spatial dimension and read as follows:
\begin{gather}
  C_e(T_e) \frac{\partial T_e}{\partial t} = \frac{\partial}{\partial x} \left( k_e(T_e,T_l) \frac{\partial T_e}{\partial x} \right) - G[T_e-T_l] + S(x,t) \label{TTM1} \\
  C_l(T_l) \frac{\partial T_l}{\partial t} = \frac{\partial}{\partial x} \left( k_l(T_e,T_l) \frac{\partial T_l}{\partial x} \right) + G[T_e-T_l] \label{TTM2}
\end{gather}
where $C_{e/l}$ and $k_{e/l}$ are the electronic/lattice heat capacity and conductivity, respectively, $G$ is the electron-phonon coupling factor and $S(x,t)$ is the heat source determined by the laser pump pulse \cite{qiu1994a,norr2003a}. The usual considerations of the TTM for elemental metals epmloy the fact that the heat is dominantly conducted by the conduction band electrons and that the phononic heat conductivity is comparably small. This allows the omission of the first term in (\ref{TTM2}) \cite{anis1974a,qiu1994a,norr2003a,chen2006a}. In SRO, however, the heat is carried by the electrons and lattice in approximately equal parts \cite{shep1998a,Maek2005} which is why we keep this term in (\ref{TTM2}).

As motivated above, the electron-electron scattering rate is very large in SRO \cite{kost1998}. The electrons are thus not able to ballistically transport energy out of the excited region into deeper parts of the sample. In addition, SRO is known to have a very fast electron-phonon relaxation time (i.e. large $G$) on the order of a few hundred femtosecond \cite{korf2008a,boja2012a} which is also much faster than any diffusion processes of electrons and phonons. When considering the structural dynamics in thin SRO films on the timescale of a few tens of picoseconds, we can thus disregard the diffusion terms in (\ref{TTM1}) and (\ref{TTM2}). This simplifies the problem considerably and the eventual expansion profile (caused by the thermal stress profile) in SRO is proportional to the exponentially decaying profile of the deposited energy density where the proportionality constant is given by the Gr\"uneisen parameter $\gamma$ \cite{ashc1976a}. The energy density profile is given by the derivative of Lambert-Beer's law and thus defined by the optical penetration depth $\xi_{\mathrm{opt}}$ of the pump laser light at 800~nm wavelength.

An implication of the very fast electron-phonon coupling in SRO is the fact that the thermal stress is built up quasi-instantaneously which launches coherent LA phonons up to very high frequencies \cite{korf2008a,boja2012a,herz2010a,herz2012b,herz2012a,herz2012c}. Similar to the temperature considerations above the structural dynamics are reduced to one spatial dimension. The coherent longitudinal (plane) strain waves traverse the excited metal layer until the entire coherent vibrational energy has left into the substrate and a quasi-statically thermally expanded layer is left. In the following sections we show that the coherent phonon dynamics inside the excited SRO thin film and the underlying STO substrate can be accurately monitored by UXRD experiments. For the later analysis we define the characteristic timescale of sound propagation through a thin film by $T_{\mathrm{sound}}=d/v_{\mathrm{sound}}$ where $d$ is the film thickness and $v_{\mathrm{sound}}$ is the longitudinal sound velocity perpendicular to the sample surface.

\section{Experimental Results} \label{ExpRes}

The UXRD experiments presented and discussed in the following were conducted on thin SRO films of different thickness epitaxially grown on a STO substrate. The samples were excited by ultrashort optical laser pulses at 800~nm wavelength and the triggered structural dynamics within the first few tens of picoseconds were observed by UXRD employing the Plasma X-Ray Source (PXS) at the University of Potsdam, Germany \cite{schi2012a}.
\begin{figure*}[tb]
  \centering
  \includegraphics[width=17.3cm]{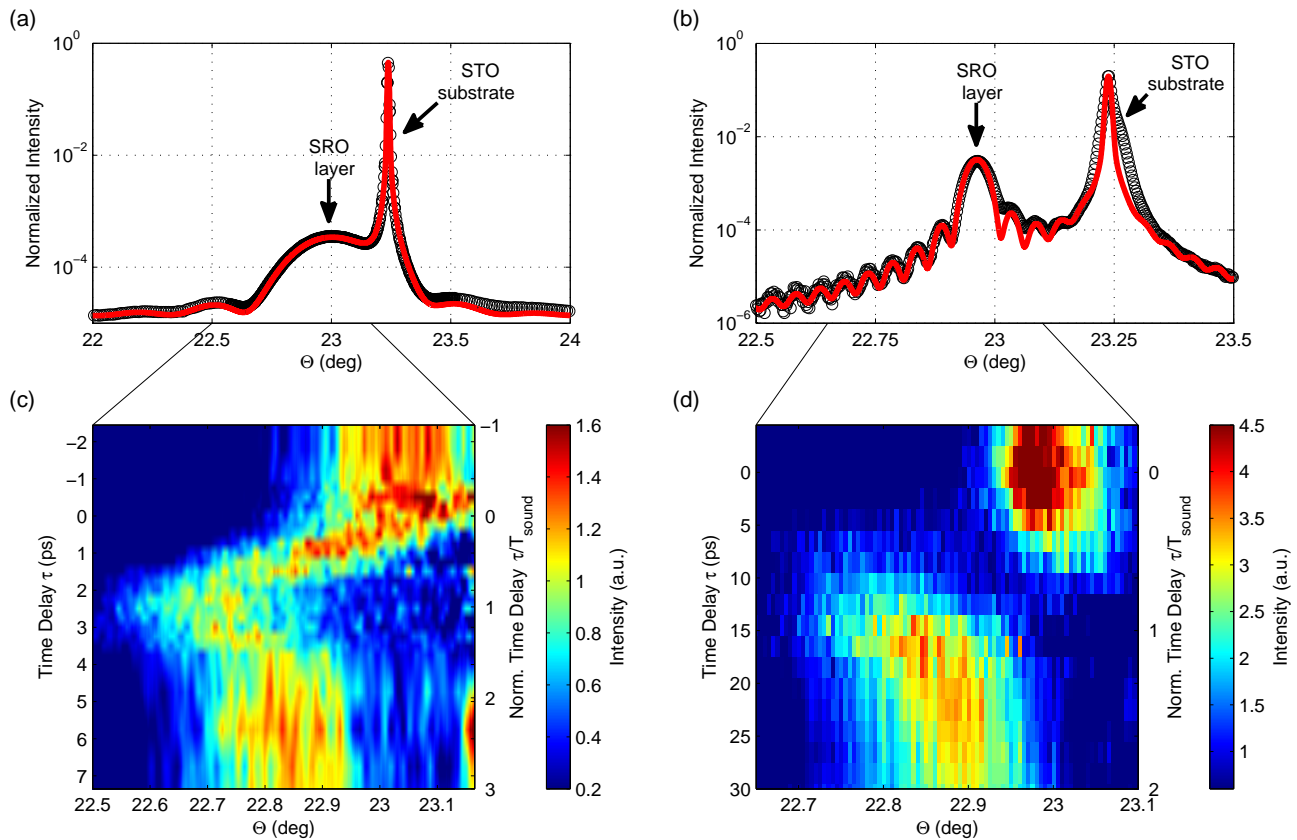}
  \caption{Static XRD curves (symbols) and dynamical XRD simulation of a (a) 15.4~nm and a (b) 94.8~nm SRO layer on STO around the (002) reflections. The data were recorded at the ESRF ($E_{\mathrm{ESRF}}=12.0$~keV) and at BESSY~II ($E_{\mathrm{BESSY}}=8.66$~keV), respectively, at x-ray photon energies differing from the characteristic Cu K$_{\alpha1}$ line ($E_{\mathrm{PXS}}=8.048$~keV) generated by the PXS. The Bragg angle axes in panel (a) and (b) were thus converted to an artificial x-ray photon energy $E_{\mathrm{PXS}}$ to be comparable to the plots (c) and (d). The lower panels present the UXRD data recorded with the PXS on the SRO Bragg peak of the (c) thinner and (d) thicker SRO thin film. The excitation fluences of the 800~nm pump pulses are 30 mJ/cm$^2$ and 20 mJ/cm$^2$, respectively.}
  \label{fig1}
\end{figure*}
This laser-based x-ray source utilizes hard x-ray pulses at the characteristic Cu K$_{\alpha1}$ and K$_{\alpha2}$ line ($E_{\mathrm{PXS}}^{(1)}=8.048$~keV and $E_{\mathrm{PXS}}^{(1)}=8.028$~keV) with a pulse duration of $\approx 200$~fs. The PXS thus provides the temporal and spatial resolution required for studying coherent acoustic phonon dynamics in thin crystalline films with a thickness below a few hundred nanometer.

The SRO thin film samples were pre-characterized by static x-ray diffraction (XRD) at synchrotron-based x-ray sources. The results of the static $\Theta$-$2\Theta$ scans around the respective (002) reflections are represented by the symbols in Fig.~\ref{fig1}(a) and (b). As expected, the thicker SRO film exhibits a narrow and intense Bragg reflection whereas the Bragg peak of the thinner SRO film is much broader and weaker. The solid lines are results of dynamical XRD simulations used for the determination of the layer thickness and $c$-axis lattice parameter. The simulations revealed a thickness $d_1=15.4$~nm and an out-of-plane lattice parameter of $c_1=3.9525$~\AA{} for the thinner SRO layer (Fig.~\ref{fig1}(a)) and $d_2=94.8$~nm and $c_2=3.9493$~\AA{} (Fig.~\ref{fig1}(b)) for the thicker layer. The different lattice parameters are consistent if one considers the epitaxy with the cubic STO substrate ($c_{\mathrm{sub}}=3.905$~\AA) and the relaxation of the substrate-induced tetragonally distorted SRO unit cell as the layer thickness increases. Employing the longitudinal sound velocity of SRO, $v_{\mathrm{sound}}=6.312$~nm/ps (Ref.~\citenum{Yama2004}), the derived layer thicknesses imply sound transit times of $T_{\mathrm{sound}}^{(1)}=2.45$~ps and $T_{\mathrm{sound}}^{(2)}=15.0$~ps for the thinner and thicker SRO layer, respectively. The two samples are chosen in order to represent the limiting cases of layers with thickness smaller and larger than the literature value of the optical penetration depth at 800~nm, $\xi_{\mathrm{opt}}^{\mathrm{lit}} \approx 52$~nm (Ref.~\citenum{kost1998}), respectively.

The UXRD data recorded at the PXS on the thinner and thicker SRO layer using a pump fluence of 30 and 20 mJ/cm$^2$ are shown in Fig.~\ref{fig1}(c) and (d), respectively. Due to the limited signal-to-noise ratio and angle-resolution of the PXS the UXRD data quality is poorer for the thinner SRO layer and the respective SRO Bragg peak of the unexcited sample is not clearly separated from the substrate peak. Nevertheless, the data can be unambiguously analyzed in terms of transient SRO Bragg peak shifts by proper substrate subtraction and thus allow for a clear understanding of the coherent and incoherent phonon dynamics in the photoexcited SRO layers.

The UXRD data exhibit transient changes of the SRO Bragg peak positions just after the laser excitation at $\tau=0$. This directly implies that the quasi-instantaneous heating of the SRO lattice triggers certain structural dynamics inside the SRO layers. At first sight, these dynamics appear to be qualitatively different. In case of the thinner SRO layer we observe a continuous shift of the Bragg peak towards lower angles for $0<\tau<T_{\mathrm{sound}}$ (region I), followed by a slight continuous backshift for $T_{\mathrm{sound}}<\tau<2T_{\mathrm{sound}}$ (region II) until it reaches a new quasi-stationary position for $\tau>2T_{\mathrm{sound}}$ (region III). This quasi-stationary expansion represents the thermal expansion of SRO due to the absorbed energy of the exciting laser pulse.
\begin{figure}[tb]
  \includegraphics[width=8.5cm]{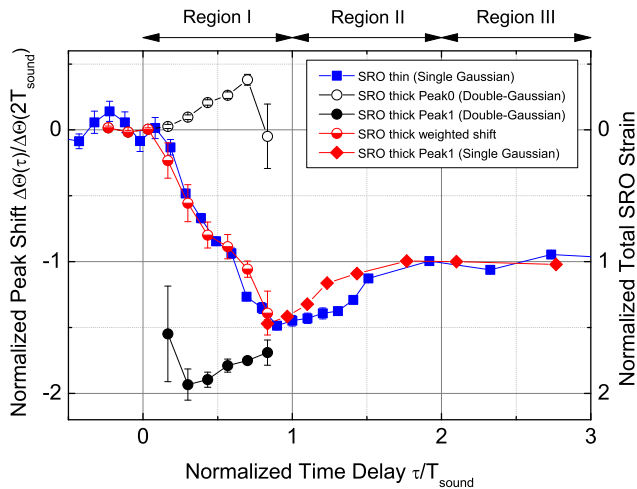}
  \caption{Normalized transient shift and splitting of the SRO Bragg peak extracted from the UXRD data shown in Fig.~\ref{fig1}(c) and (d). The continuous shift observed in the thinner SRO layer is represented by the blue squares (single-Gaussian fitting). The empty and solid black bullets show the transient Bragg angles of the decaying initial and the rising shifted peak for the thicker SRO layer (double-Gaussian fitting). For $\tau/T_{\mathrm{sound}} > 1$ the red diamonds indicate the continuous backshift of the thicker SRO layer (single-Gaussian fitting). The intensity-weighted shift for the thin SRO layer is shown by the red half-filled bullets. The error bars represent the 68\% confidence interval of the fitting parameter and lines are a guide to the eye.}
  \label{fig2}
\end{figure}
In contrast, the thicker SRO layer does not exhibit a continuous shift for $0<\tau<T_{\mathrm{sound}}$ (region I). Instead, we find a splitting of the initial Bragg peak into two distinct reflections at intermediate times until the initial peak has disappeared. Similar Bragg peak splittings have been previously observed in photoexcited bulk crystals and thin films \cite{Rose1999a,lee2008a,quir2012a}, however, either the signatures were relatively weak compared to the bulk reflection or a thorough description of the underlying structural dynamics is missing. The comparison of the experimental results in Fig.~\ref{fig1}(c) and (d) rises the question whether the structural dynamics responsible for the observed features are indeed qualitatively different. This issue is addressed in the subsequent sections.

We extracted the transient Bragg angles of the measured SRO peaks by fitting the data with single and double-Gaussian functions. The obtained peak positions are displayed in Fig.~\ref{fig2} where we plot the Bragg angle change normalized to the quasi-stationary value after $2T_{\mathrm{sound}}$, $\Delta\Theta(\tau)/\Delta\Theta(2T_{\mathrm{sound}})$, versus time delay in units of $T_{\mathrm{sound}}$. Since the angle changes are fairly small they are proportional to the average strain variations of the SRO layer. This is indicated by the secondary y-axis in Fig.~\ref{fig2} presenting the normalized total SRO strain. The plot verifies the features visible in the contour plots in Fig.~\ref{fig1}(c) and (d). The thinner sample exhibits a continuous shift of a single SRO peak (blue squares). In contrast, the SRO peak in the thicker sample for normalized time delays up to $T_{\mathrm{sound}}$ (region I) shows a peculiar splitting. The Bragg angles of the initial and displaced peaks are represented by the black empty and solid bullets in Fig.~\ref{fig2}, respectively. In addition, we observe that both individual peaks shift to higher Bragg angles as they decay and rise. Despite the unequal behaviour of the respective SRO peaks in region I, the normalized transient shifts in region II and III ($\tau>T_{\mathrm{sound}}$) appear to be comparable in both SRO layers. In addition, we evaluate the intensity-weighted transient shift of the thicker SRO layer in region I which is shown by the red half-filled bullets in Fig.~\ref{fig2}. This curve represents the sum of the individual shifts weighted by the intensity of the respective Bragg peak at each time step. We find that this curve coincides with the transient peak shift of the thinner SRO layer. This immediately implies that the time-evolution of the average strain is identical in both layers as will be discussed below in more detail.

\section{Discussion}

\subsection{Splitting versus Shifting of Bragg Peaks} \label{SplitVsShift}

In this section we focus on the time region I ($0<\tau<T_{\mathrm{sound}}$) where we observe a splitting of the SRO Bragg peak for the thicker sample. We perform numerical lattice-dynamics calculations and use the results in order to simulate the transient XRD response of the SRO layer.

It has recently been shown that a linear-chain model of masses and springs is well-suited to calculate the lattice dynamics triggered by the quasi-instantaneous thermal stress in SRO \cite{herz2012b}. Moreover, the results of these calculations can be easily used to accurately simulate the transient x-ray response of such photoexcited nanolayered samples employing dynamical XRD theory. \cite{herz2012a}. We therefore apply this toolbox to the present case of laser-excited SRO layers of different thickness in order to elucidate the general features of the structural dynamics.

The details of the linear-chain model simulations for the calculation of the photoinduced lattice dynamics of a thin SRO film on a STO substrate are given elsewhere \cite{herz2012b}. The essential ingredient is the assumption of an instantaneous rise of an isotropic thermal stress in SRO due to incoherently excited phonons at $\tau=0$. In a recent UXRD study on hexagonal LuMnO$_3$ anisotropic elastic properties had to be accounted for which are due to the lower crystal symmetry compared to SRO \cite{lee2008a}. However, the SRO unit cell only slightly deviates from cubic symmetry (pseudocubic) implying rather isotropic elastic properties \cite{Yama2004}. The fact that the thermal stress in SRO builds up quasi-instantaneously has been proven to be valid by several UXRD experiments on a few-ps timescale \cite{herz2012b,herz2012a,herz2012c}. In fact, very tiny phase shifts of photoexcited coherent phonons modes in superlattices have been observed which evidence a crossover from a finite rise time ($\approx 200$~fs) to an instantaneous onset of the displacive thermal stress \cite{boja2012a}. Nonetheless, as we discuss below, the assumption of an instantaneous driving force is sufficiently good for the dynamics considered in this report. In Section~\ref{TTM} we mention that the very fast electron-phonon relaxation in SRO implies a thermal stress profile given by the exponential absorption of the pump light. Accordingly, we start our calculations by assuming an exponential thermal stress profile with a $1/e$ decay length defined by the optical penetration depth of SRO, $\xi_{\mathrm{th}}=\xi_{\mathrm{opt}}^{\mathrm{lit}}=52$~nm \cite{kost1998}.

\begin{figure}[tb]
  \includegraphics[width=8.5cm]{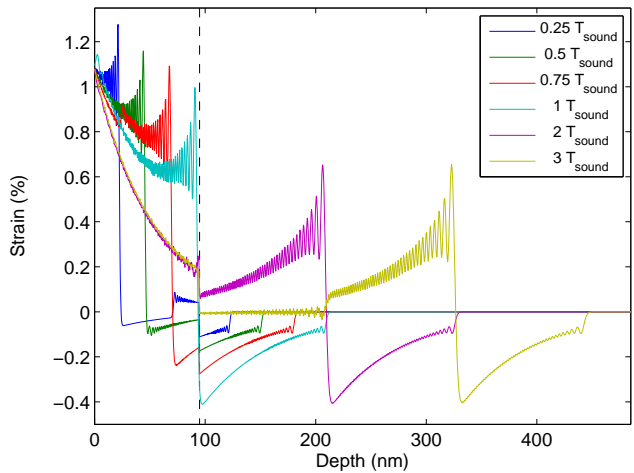}
  \caption{Calculated photoinduced strain field of the thicker SRO layer on STO for selected time delays. The laser excitation at $\tau=0$ launches coherent LA phonons which form sharp strain waves propagating through the layer and the substrate.}
  \label{fig3}
\end{figure}

The spatio-temporal strain field for the thicker SRO layer on STO resulting from the lattice dynamics calculations is shown in Fig.~\ref{fig3}.
The laser-induced thermal stress sets in at $\tau=0$. The instantaneous rise of this driving force coherently excites LA phonon modes up to very high frequencies. In total, the superposition of all coherent phonons results in propagating strain wave fronts starting at the air-SRO and SRO-STO interfaces where the thermal stress is not balanced and exhibits large gradients. In case of the 94.8~nm SRO layer whose thickness is almost twice the optical penetration depth of 800~nm light, $\xi_{\mathrm{opt}}^{\mathrm{lit}}$, the dominating strain front is an expansion wave launched at the surface. The expansion wave starting at the SRO-STO interface has much lower amplitude but is still visible (blue line in Fig.~\ref{fig3}). Thus the essential feature inferred from the calculations for the thicker SRO layer is the generation of a propagating wave front dividing the SRO layer into an expanded sublayer near the surface whereas the remaining sublayer gets slightly compressed in total for $0<\tau<T_{\mathrm{sound}}$ (green and red line in Fig.~\ref{fig3}). The thickness of these sublayers is gradually increasing and decreasing with time, respectively. The slight compression of the decreasing sublayer is addressed in more detail below. At later times the coherent strain waves have propagated into the substrate forming a bipolar strain pulse \cite{thom1986a} leaving a stationarily expanded SRO layer. A very small acoustic mismatch of SRO and STO results in a negligible reflection coefficient of acoustic waves at the SRO-STO interface \cite{land1987a}. That is, the strain waves do not travel back and forth several times inside the SRO layer which would lead to a breathing mode of this layer as observed in other material combinations or in free-standing films \cite{nico2011a,Li2009}.

A closer look at the transient strain fields in Fig.~\ref{fig3} reveals that the propagating wavepackets are superimposed by a peculiar fine structure which is most pronounced for the tensile component of the bipolar strain pulse. Since we solve the differential equations of the linear chain analytically these features are no numerical artifacts \cite{herz2012b}. In fact, these high-frequency oscillations are a characteristic feature of a discretized linear chain and do not occur in elastic continuum models \cite{thom1986a}. They are a result of the fact that the motion on the linear chain is essentially initiated at the surface and interface due to the large gradients of the thermal stress at these points. In addition to the initial displacement of the outermost masses on the linear chain an oscillatory motion of these is launched\footnote{Note, that the frequency of these oscillations actually depends on the particular discretization of the linear chain. The presented simulations assume one mass on the linear chain per unit cell of SRO and STO. If one refines the linear chain to include one mass per lattice plane (i.e. two masses per perovskite unit cell along [001]) the frequency will double.}. A thorough description of this high-frequency component is out of the scope of this report and shall be given elsewhere.
Note that these high-frequency modes require a very fast build-up of the thermal stress to be efficiently excited, i.e. a sufficiently short pump pulse and a fast electron-phonon relaxation. However, in real SRO and STO crystals at room temperature the lifetime of acoustic phonons of such high frequency is expected to be very short due to anharmonic phonon-phonon scattering \cite{kore2006a,klie2011a,herz2012c}.

\begin{figure}[tb]
  \includegraphics[width=8.5cm]{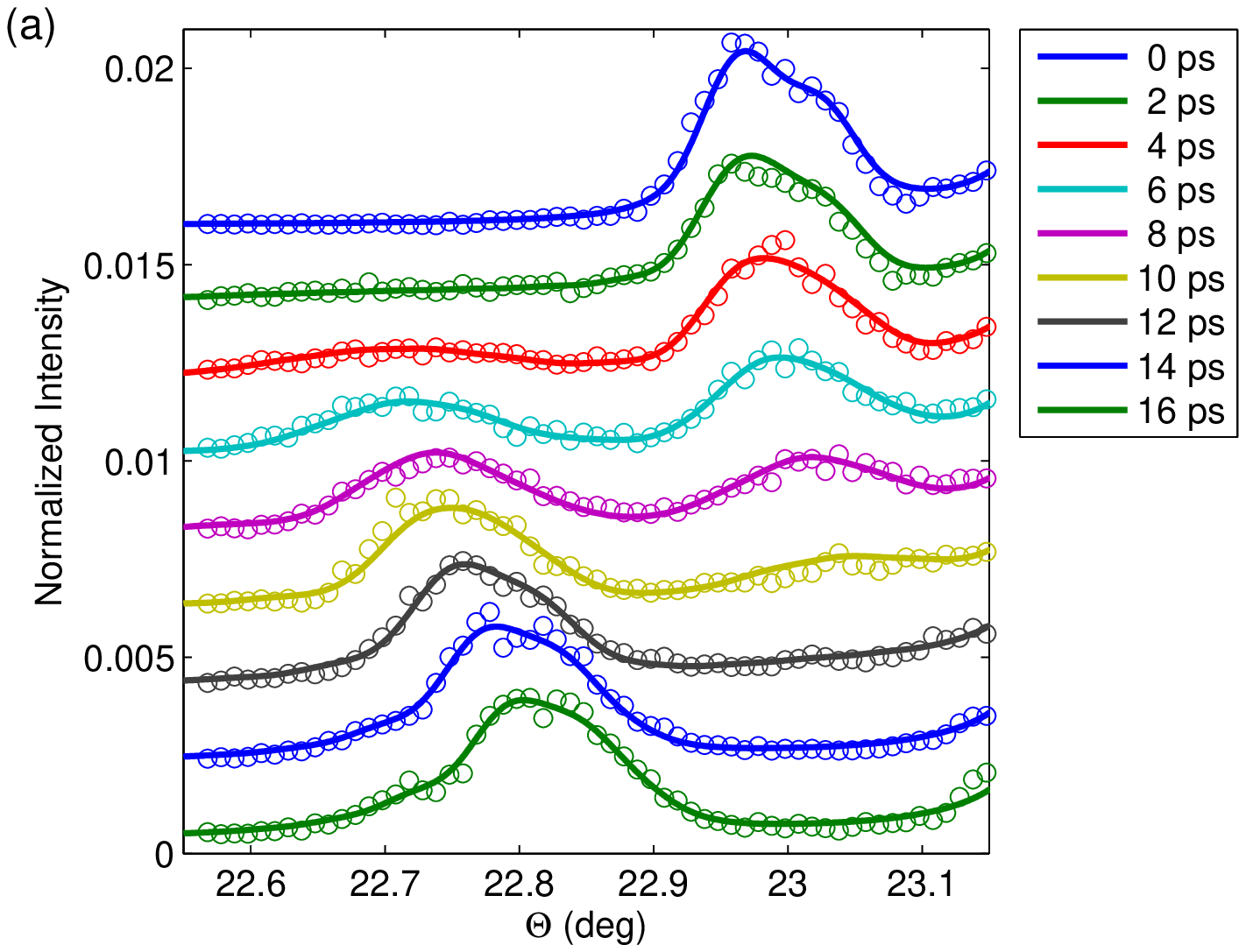} \\
  \includegraphics[width=8.5cm]{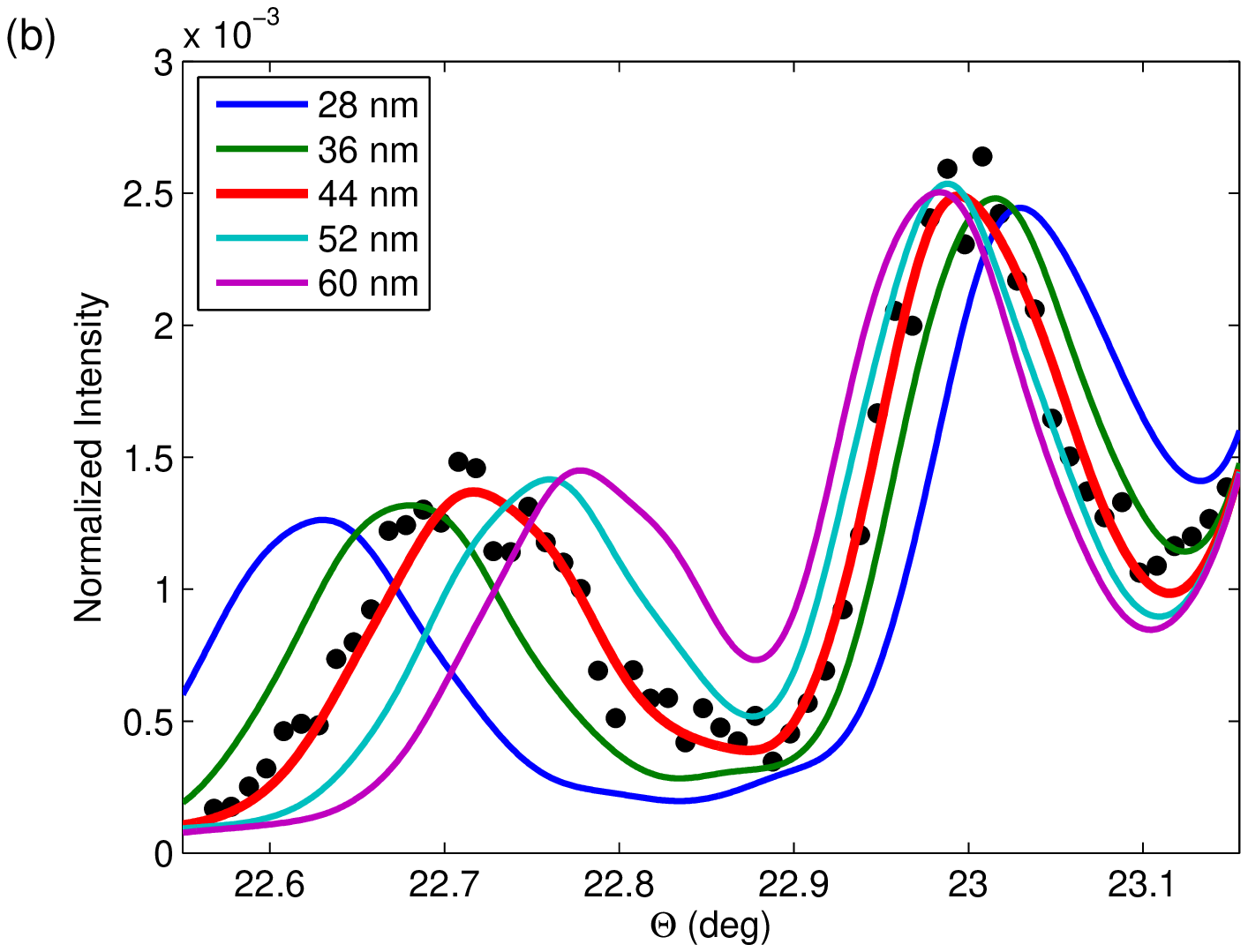}
  \caption{(a) Comparison of the transient diffraction curves measured on the thicker SRO layer at the PXS (symbols) with simulations (solid lines). The experimental data were recorded at a pump fluence of 30~mJ/cm$^2$. Time delay increases from top to bottom and the curves are displaced for clarity. The simulations employ an exponential thermal stress profile with decay length $\xi_{\mathrm{th}}=44~\mathrm{nm}<\xi_{\mathrm{opt}}^{\mathrm{lit}}$. (b) Transient diffraction curve at $\tau=6$~ps (symbols) and simulations assuming different $\xi_{\mathrm{th}}$.}
  \label{fig4}
\end{figure}

In order to correlate the identified transient features of the photoinduced lattice dynamics to the transient signals in UXRD experiments we calculate the diffraction curves of the sample at each time step utilizing the transient lattice deformations presented in Fig.~\ref{fig3}. Figure~\ref{fig4}(a) compares the numerical results with experimental data similar to the data shown in Fig.~\ref{fig1}(d) but at a higher pump fluence of 30~mJ/cm$^2$ and without substrate subtraction. The presented simulation accounts for the instrument function of the PXS and assumes a decay length of the thermal stress profile of $\xi_{\mathrm{th}}=44$~nm to obtain a good match. This deviation from the optical penetration length $\xi_{\mathrm{opt}}^{\mathrm{lit}}$ is addressed below.
The simulation reproduces the UXRD data very precisely. In particular, the splitting of the SRO Bragg peak and the individual shifts as the peaks decay and grow are in perfect agreement. We can now correlate the features revealed by the UXRD data to the structural dynamics obtained from the numerical lattice-dynamics calculations (Fig.~\ref{fig3}). The initial rocking curve at $\tau=0$ represents the unexcited SRO layer. As discussed above, the photoinduced thermal stress essentially launches a strain wave front at the surface which generates a gradually growing expanded sublayer. This rather thin expanded layer gives rise to the appearance of the broad and weak extra peak at lower angles. The peak grows as time delay increases since the thickness of this expanded sublayer increases. Simultaneously, the slightly negatively strained sublayer decreases in thickness leading to the gradual disappearance and broadening of the initial SRO peak. This observation immediately verifies the very fast generation of the thermal stress which is required for the sharp division into differently strained sublayers\footnote{The clear splitting implies a rise time of the thermal stress much faster than the characteristic timescale of layer expansion, $T_{\mathrm{sound}}$. However, an exact determination of the rise time is not possible with such a sample structure (cf. Ref.~\citenum{boja2012a})}.

The UXRD data also show a slight shifting of the individual peaks to higher angles as they rise and fall. As evidenced in Fig.~\ref{fig4}(a), the simulation also accurately reproduces this behaviour. The reason of the transient shifts of both peaks can also be found in the calculated lattice dynamics and is related to the exponential profile of the thermal stress and of the resulting layer expansion. As can be inferred from Fig.~\ref{fig3}, the expansion wave launched at the surface gradually imprints an exponential expansion profile into the SRO film. The average strain of the expanded sublayer thus decreases as time increases. Accordingly, the angular displacement of the related rising Bragg peak is largest just after time zero and slowly decreases as time increases.
The reason for the slight shift of the initial peak is a little more subtle. In simple words one may say that at early times the strongly expanding near-surface region of the SRO layer increasingly squeezes the less expanding near-interface region until the entire layer is expanded. In general, a gradient in the (instantaneous) thermal stress profile launches acoustic sound waves. In the considered case, the most dominant gradients appear at the surface and interface where the sharp strain waves are triggered. However, also the exponentially varying negative gradient inside the SRO film is responsible for a compressive strain wave component propagating towards the substrate. This compressive strain component is dominant over the tiny tensile strain wave from the SRO-STO interface (green and red line in Fig.~\ref{fig3}). Hence it causes an increasing compression of the corresponding sublayer as time increases which is represented by the slight shift of the initial SRO Bragg peak towards higher angles. At later times it turns out that the exponential gradient is also responsible for the corresponding shape of the bipolar strain pulse in the substrate  (magenta and yellow line in Fig.~\ref{fig3}).

The above explanations rely on the exponential dependence of the thermal stress and thus expansion profile which suggest that the particular shifting behaviour of the two peaks is sensitive to, e.g., the decay length $\xi_{\mathrm{th}}$ of these profiles. To verify this we performed simulations with different values of this decay length. The final strain of the SRO layer and thus the Bragg angle of the SRO peak at late times is held constant. A comparison of the transient rocking curve at $\tau=6$~ps with these simulations is shown in Fig.~\ref{fig4}(b). Indeed, the shorter the decay length the more shifted is the rising (decaying) peak due to a relatively larger average expansion (compression) of the near-surface (near-interface) sublayer at early times. Surprisingly, the best fit of the UXRD data shown in Fig.~\ref{fig4}(a) is achieved assuming a decay length of $\xi_{\mathrm{th}}=44$~nm. As discussed in Section \ref{TTM}, the extraordinary fast electron-phonon relaxation in SRO should generate a thermal stress and expansion profile having a decay length $\xi_{\mathrm{th}}$ equal to the optical absorption length $\xi_{\mathrm{opt}}^{\mathrm{lit}}$. However, the decay length deduced from the UXRD data is significantly lower than the expected 52~nm. Even if one allowed for any ballistic or diffusive thermal transport before the energy is coupled into the lattice this result could not be explained since these processes tend to flatten out any gradients resulting in a longer decay length.

There are two effects which could cause the unexpected steepness of the observed expansion profile. First, the optical penetration depth could possibly be different in the considered thin films as compared to the bulk material due to finite size effects and/or variations of the optical constants by stationary strains that are induced by epitaxy \cite{liu2006,veis2009a}. We performed spectroscopic ellipsometry measurements on the thick SRO layer and found an optical penetration depth of $\xi_{\mathrm{opt}}^{\mathrm{exp}} = 48$~nm at a wavelength of 800~nm. This value is indeed slightly smaller than the literature bulk value of 52~nm \cite{kost1998} but still significantly larger than the lengthscale of the expansion profile deduced from the UXRD data.

The second possible effect which could cause a steeper expansion profile is a weak temperature dependence of the Gr\"uneisen parameter $\gamma$ describing the ratio of thermal expansion and deposited energy. In general, this parameter is nearly material-independent and shows almost no temperature dependence, however, a slight increase of $\gamma$ with temperature can be observed in several materials \cite{kitt1996a}. An increase of $\gamma$ with $T$ would cause a larger expansion of the near-surface regions of the SRO layer relative to the deposited energy density profile which would result in a steeper expansion profile as revealed by the UXRD data.

In the following we briefly turn to the lattice dynamics of the thinner SRO layer whose thickness is much smaller than the optical penetration length of the 800~nm pump light. Here, we can extend the concept of differently strained sublayers. Since the deposited energy density is comparable near the surface and interface, respectively, the launched strain wave fronts are also similar in amplitude. Accordingly, one finds three sublayers of different strains which should in principle cause a more complicated splitting of the layer Bragg peak. However, due to the small thickness of the SRO layer the corresponding Bragg peak is very broad and thus prevents any splitting from being observed. The required strain amplitudes to observe a splitting for the very thin film would be too large. This complex splitting of Bragg peaks could possibly be visible on thicker layers of materials having an accordingly larger optical penetration depth.

\subsection{Universal Features of Lattice Dynamics in SRO Thin Films}

In the previous section we discussed the qualitatively different UXRD signatures while the coherent strain waves pass through the layer once ($0<\tau<T_{\mathrm{sound}}$). As already mentioned in section \ref{ExpRes}, the transient shifts appear very similar for the two thin films with different thicknesses at times later than $T_{\mathrm{sound}}$ (region II and III). Moreover, the transient weighted shifts (average strain) are almost identical for all times. In the following we thus address the evolution of the average layer strain and how it depends on the ratio of the layer thickness $d$ and the decay length of the thermal stress $\xi_{\mathrm{th}}$.

\begin{figure}[tb]
  \includegraphics[width=8.5cm]{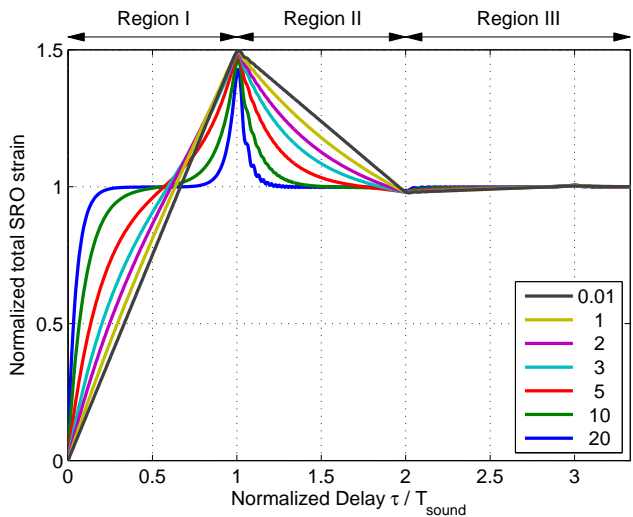}
  \caption{Transient average SRO strain for different ratios $d/\xi_{\mathrm{th}}$.}
  \label{fig5}
\end{figure}

In order to investigate the effect of a varying ratio $d/\xi_{\mathrm{th}}$, we calculate the spatio-temporal strain fields for the thick SRO layer ($d$ fixed) at various values of $\xi_{\mathrm{th}}$. We then extract the transient average SRO strain normalized to the final strain at $\tau>2T_{\mathrm{sound}}$ and plot the results versus normalized time delay $\tau/T_{\mathrm{sound}}$ in Fig.~\ref{fig5}.
We identify three regions of qualitatively different lattice dynamics which, however, show universal behaviour independent of the ratio $d/\xi_{\mathrm{th}}$. In region I ($0<\tau<T_{\mathrm{sound}}$) the coherent strain waves cause the average SRO strain to monotonically increase to 150\% of the final thermal expansion. This value of maximum strain does not depend on the ratio $d/\xi_{\mathrm{th}}$, however, the precise evolution of the total SRO strain is influenced by the ratio. It changes from linear to jump-like as the ratio decreases. Similar observations hold for region II ($T_{\mathrm{sound}}<\tau<2T_{\mathrm{sound}}$) but here we find a monotonically decreasing strain until it reaches the final value determined by the pure thermal expansion due to the increased SRO temperature. In region III ($\tau>2T_{\mathrm{sound}}$) the SRO strain remains quasi-constant until it relaxes back to zero via heat diffusion into the substrate on a nanosecond timescale (not included in the model and not shown).

The quantitative difference of the total SRO strain evolution is simply explained by the different shapes of the strain waves launched at the surface and/or interface. In case of $d/\xi_{\mathrm{th}} \ll 1$ (grey line in Fig.~\ref{fig5}) square-like strain pulses are launched at the surface and interface with comparable amplitudes which results in the piecewise linear evolution of the total strain. In contrast, if $d/\xi_{\mathrm{th}} \gg 1$ (blue line in Fig.~\ref{fig5}) only a small portion underneath the surface of the SRO layer is excited which then rapidly expands just after time zero. Moreover, a rather localized bipolar strain pulse is launched \cite{thom1986a} which has a vanishing integral strain as we discuss below. Around $\tau=T_{\mathrm{sound}}$ the bipolar strain pulse leaves the SRO layer and causes the spike-like transient increase of the total SRO strain. Altogether, Figure~\ref{fig5} evidences that the calculations precisely predict the average SRO strain dynamics derived from the transient (weighted) shift of the SRO Bragg peaks shown in Fig.~\ref{fig2}. In particular, the slight differences of the red and blue data in Fig.~\ref{fig2} (SRO layers of different thickness) can thus mainly be attributed to the different ratios $d/\xi_{\mathrm{th}}$.

\begin{figure}[tb]
  \includegraphics[width=8.5cm]{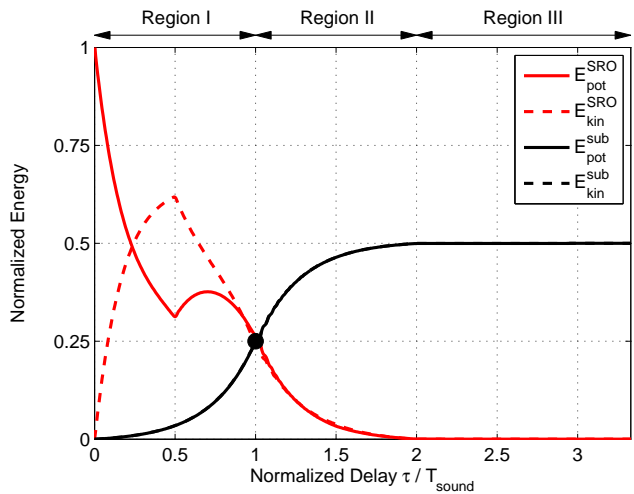}
  \caption{Transient normalized potential (solid lines) and kinetic energy (dashed lines) of the coherent strain waves inside the SRO layer (red) and the STO substrate (black). The energies were derived from the calculation presented in Fig.~\ref{fig3}. Note that the potential and kinetic energy in the substrate are identical and thus the respective curves are overlayed. The black bullet indicates the point at which all normalized energies are identical. This point is independent of $\xi_{\mathrm{th}}$}
  \label{fig6}
\end{figure}

The striking feature in Fig.~\ref{fig5} is that the normalized total strain in the SRO layer at $\tau=T_{\mathrm{sound}}$ and for $\tau>2T_{\mathrm{sound}}$ (region III) is independent of the ratio $d/\xi_{\mathrm{th}}$. In particular, the maximum strain due to the coherent phonon dynamics is always 50\% larger than the steady-state strain after $2T_{\mathrm{sound}}$ due to incoherent phonons (i.e. heat). To give an explanation we employ the potential and kinetic energy of the coherent strain waves inside the SRO layer and the STO substrate, respectively, which are shown in Fig.~\ref{fig6}.
At $\tau = 0$ the thermal stress is generated quasi-instantaneously by the absorption of the laser pulse which increases the equilibrium distance of atoms in the lattice. Therefore, the thermal stress gives rise to an initial potential energy since the SRO layer is now compressed relative to the new equilibrium state. The kinetic energy is zero since no atomic motion has started at this point. As explained in section \ref{SplitVsShift}, the thermal stress launches propagating strain waves inside the SRO layer which is evidenced by the increasing kinetic and decreasing potential energy of SRO. The increase in both kinetic and potential energy inside the STO substrate is due to the corresponding strain pulse launched from the interface into the substrate (cf. Fig.~\ref{fig3}). At $\tau=T_{\mathrm{sound}}$ all energies are identical and the key point is that this observation is independent of the ratio $d/\xi_{\mathrm{th}}$ (not shown) which is indicated by the black bullet. In particular, the substrate-related energies reached half of their maximum values since the compressive part of the bipolar strain pulse has entered the substrate while the tensile part still remains in the SRO layer. This implies that the integral strain of the two individual parts of the bipolar strain pulse (tensile and compressive) is equal but has opposite signs. In contrast to the initially ($\tau=0$) compressed state of the SRO layer relative to the new equilibrium state after $2T_{\mathrm{sound}}$, the coherent superposition of the longitudinal strain waves at $\tau=T_{\mathrm{sound}}$ results in a tensile state relative to the new equilibrium (cf. Fig.~\ref{fig3}). At this point in time the potential energy of the SRO layer dropped to 1/4 of its initial value. Since the potential energy generally is proportional to the square of the strain we conclude that the relative expansion at $\tau=T_{\mathrm{sound}}$ is 1/2 of the initial relative compression. Hence the maximum absolute strain relative to the unexcited state at $\tau=T_{\mathrm{sound}}$ has to be 150\% of the equilibrium (thermal) strain after $2T_{\mathrm{sound}}$.

As mentioned in section \ref{SplitVsShift} the perfect acoustic matching of SRO and STO prevents the coherent strain waves from being (partially) reflected back into the SRO layer. If there is a significant acoustic mismatch the strain waves would travel back and forth until all vibrational energy has been transferred to the substrate. This would result in an oscillatory behaviour of the total SRO strain as was observed by UXRD in a photoexcited Gold layer on a Mica substrate \cite{nico2011a}. The limiting case of such an acoustic breathing of a metal film is represented by free-standing films where the amplitude decay of the acoustic waves is merely given by internal damping effects. Such breathing of free-standing Al films has been studied by femtosecond electron diffraction \cite{Li2009}. Ideally, the maximum strain at $\tau=T_{\mathrm{sound}}$ due to the coherent structural dynamics in an acoustically decoupled photoexcited thin film is 100\% above the steady-state thermal expansion which is related to the displacive nature of the excitation mechanism \cite{barg2004b,barg2006a}. Any type of energy loss---be it coherent by transmission losses into the substrate or incoherent by internal damping and/or scattering---leads to a decrease of the maximum coherent strain at $\tau=T_{\mathrm{sound}}$. In the present case of a semi-infinite substrate with perfect acoustic matching the coherent strain maximum of the metal layer is reduced by a factor of 2 since a part of the initial potential energy is transferred to the substrate in form of the compressive half of a bipolar strain pulse.

\section{Conclusion}

In this report we address the issue of the one-dimensional coherent structural dynamics in thin films of metallic SrRuO$_3$ (SRO) on a dielectric substrate SrTiO$_3$ which are triggered by the absorption of ultrafast optical laser pulses. We show experimental results of UXRD experiments which probe the photoexcited coherent lattice dynamics of two SRO layers with different thickness. The observed changes of the respective SRO Bragg peaks exhibit qualitatively different features. In particular, the Bragg peak of the thicker SRO layer shows a transient splitting into two separated reflections which evidences the propagation of sharp longitudinal strain waves through the metal layer. These coherent wavepackets are caused by the exceptionally fast electron-phonon relaxation previously identified in SRO \cite{korf2008a,boja2012a}. Numerical models accounting for the photoinduced structural dynamics and dynamical XRD show excellent agreement assuming a surprisingly small decay length of the thermal stress profile of 44~nm. This deviation from the measured optical penetration depth in these thin films may be attributed to a temperature-dependent Gr\"uneisen parameter. Finally, we analyze the UXRD-calibrated coherent phonon dynamics in detail using the numerical simulations. The UXRD features can unambiguously be related to the precise structural dynamics. We discuss the features specific to the layers of different thickness (Bragg peak splitting versus continuous shift) and identify a universal evolution of the total strain of a photoexcited thin metal film of arbitrary thickness on a semi-infinite substrate. Our work gives a very precise and UXRD calibrated picture of the acoustic deformation of laser-heated metal layers on a supporting substrate and carefully relates the transient structural features to UXRD signatures. We believe that this work is very valuable for the quantitative interpretation of time-resolved scattering experiments on the complex photoinduced structural dynamics of crystalline nanolayered structures such as thin films, multilayers and superlattices.

We thank the BMBF for funding the project via grant No. 05K10IP1 and the DFG via grant No. BA2281/3-1.

\bibliography{2012_PRB_Thin_Film_Lattice_Dynamics}

\end{document}